\documentclass[twocolumn]{revtex4-1}

\usepackage{graphicx}

\begin{document}

\title{Identification of High-Momentum Top Quarks, Higgs Bosons, and $W$ 
       and $Z$ Bosons Using Boosted Event Shapes}

\author{J. S. Conway}
\author{R. Bhaskar}
\author{R. D. Erbacher}
\author{J. Pilot}
\affiliation{University of California, Davis}

\date{Oct. 10, 2016}

\begin{abstract}
At the Large Hadron Collider, numerous physics processes expected
within the standard model and theories beyond it give rise to very
high momentum particles decaying to multihadronic final states.
Development of algorithms for efficient identification of such
``boosted'' particles while rejecting the background from multihadron
jets from light quarks and gluons can greatly aid in the
sensitivity of measurements and new particle searches.  This paper
presents a new method for identifying boosted high-mass particles
using event shapes in Lorentz-boosted reference frames.  Variables
calculated in these frames for multihadronic jets can then be used as
input to a large artificial neural network to discriminate their
origin.
\end{abstract}

\maketitle

\section{Motivation}

The center-of-mass energy of $pp$ collisions at the Large Hadron
Collider (LHC) has now reached 13 TeV in 2015 and 2016 running, up
from 7 TeV and 8 TeV in the 2011 and 2012 running periods,
respectively.  The large LHC general-purpose experimental
collaborations ATLAS and CMS, working with theoretical colleagues,
have developed powerful new analysis techniques to efficiently
identify high-momentum (``highly boosted'') hadronically decaying top
quarks, Higgs bosons, and $W$ and $Z$ bosons, while rejecting the
background from multijet final states produced in QCD processes.  As
energies increase, the rate of production of highly boosted heavy
particles increases dramatically due to the increased parton-parton
luminosity at large invariant masses.  Numerous production mechanisms
for highly boosted particles exist within the standard model (SM) and
are predicted in theories beyond the standard model (BSM).

\begin{figure*}
  \begin{center}
      \includegraphics[width=\textwidth]{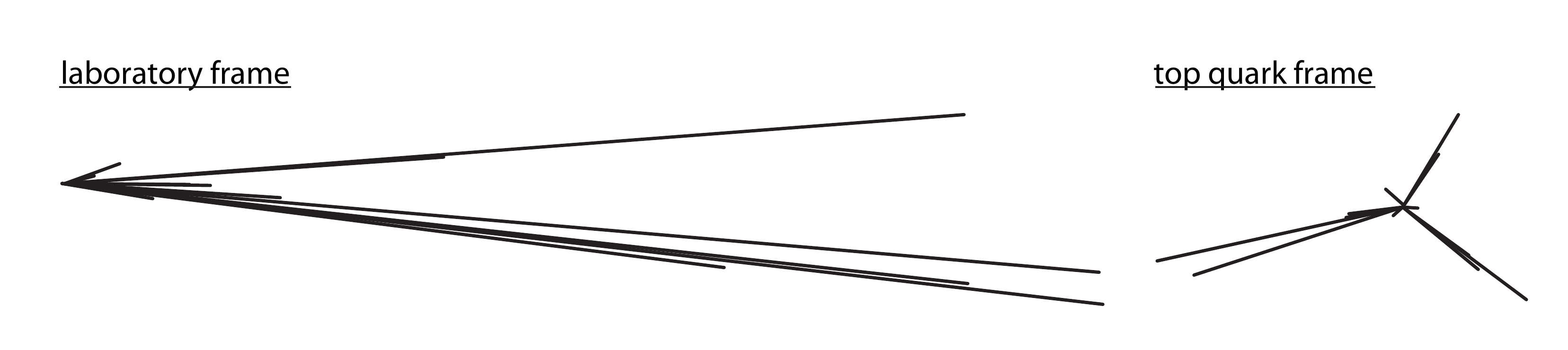}
  \end{center}
  \caption{Momentum vectors of the final state particles from top quark
           decay to three jets in the laboratory frame and in a boosted frame 
           corresponding to a top quark with the original jet laboratory momentum.}
  \label{fig-vec}
\end{figure*}

Up to now, the primary method for identifying, for example, decays of
boosted top quarks decaying to a b quark and $W$ boson, where the W
boson decays to a quark pair, has been to use the pattern of energy
(or momentum) flow in the two-dimensional space of pseudorapidity
($\eta$) and azimuthal angle ($\phi$) within a ``fat'' jet
reconstructed from observed final state particles~\cite{alg1,alg2,alg3,alg4,alg5}.
By ``fat'' we mean here a jet reconstructed using a large value of the
cone size parameter governing the jet finding algorithm, for example
the anti-$k_T$ algorithm~\cite{anti-kt} with a cone size paramter of
0.8.  The algorithms look for substructure within such jets, where
there are distinct and separable deposits of energy from the
individual jets formed from the initial-state $b$ quark and light
quark jets from the top decay.  Similar algorithms can be used to
identify boosted $W$ and $Z$ bosons decaying to quark pairs, and Higgs
bosons decaying to $b\bar{b}$.  

We present here a new approach in the identification of boosted
massive particles.  The goal is a generic algorithm which can be used
to determine the origin of individual ``fat'' jets with large
transverse energy ($E_T$) produced in $pp$ collisions at the LHC.  In
the ideal case such an algorithm will yield estimators allowing
analysts to set operating points in a multidimensional output space
and optimize the efficiency for identifying hadronically decaying $t$
quarks, $H$ bosons, and $W$ and $Z$ bosons, discriminating against jets
from light quarks and gluons.

\section{Boosted Event Shapes}

In this new algorithm, one begins with a list of the fat jet
constituents, which are reconstructed from charged particle tracks,
calorimeter energy deposits, and information from the outer muon
system.  For example the CMS experiment utilizes a sophisticated
algorithm known as ``particle flow'' (PF)~\cite{particle-flow} to
arrive at a list of particle candidates from which jets are found
using jet clustering algorithms.  Typically the PF algorithm can
identify 80-90\% of the particles in a jet, leaving relatively small
corrections to determine the overall jet energy and momentum.

Given the list of fat jet constituents, then, we successively
Lorentz-boost all of them into hypothetical reference frames
corresponding to our heavy particle ($t/H/W/Z$) masses, along the jet
momentum direction, and assuming the momentum of the heavy particle is
equal to the observed jet momentum.  In the correct reference frame
one would expect to see back-to-back jets for a $W$, $Z$, or $H$
event, and three jets in a plane for a $t$ quark decay.
Figure~\ref{fig-vec} shows the momentum vectors of the final state
particles from a top quark decay in the laboratory frame and in a
boosted frame corresponding to that of a top quark with the momentum
of the jet.  As the figure shows, one clearly sees a three-jet
structure in the boosted frame, which in the laboratory appears to
have a two-jet substructure.  This illustrates that this method has
the potential to resolve jets which strongly overlap in the laboratory
frame momentum space, and thus the detector itself.

Within each reference frame we then calculate event shape parameters
and mometum balance estimators.  These include Fox-Wolfram
moments~\cite{Fox-Wolfram}, the eigenvalues of the sphericity
tensor~\cite{sphericity}, and thrust~\cite{thrust}.  We also run the
anti-$k_T$ jet finding algorithm in the boosted frame, relative to the
boost axis and using a cone size parameter of 0.5. We then calculate
the $E_T$ of the four leading jets, the masses of pairs of the three
leading jets, and the invariant mass of the four leading jets.  If we
have boosted into the correct frame, corresponding to the true origin
of the fat jet, then we would expect that the overall momentum of the
boosted constituents should be near zero, either in terms of the total
momentum or the momentum along the boost direction. Hence we also
calculate the magnitude of the total momentum of the four leading jets
and the ratio of their longitudinal momentum sum to the total.
Generically we call these quantities ``boosted event shapes'' (BES).
This approach is inspired by earlier work at LEP aimed at
discriminating $b\bar{b}$ decays of the $Z$ boson using the ``boosted
sphericity product''~\cite{BSP}.  It differs from the approach
described in \cite{Chen}, and employed by the ATLAS experiment in
their measurement of high-$p_T$ vector boson pair prodiction cross
section~\cite{ATLAS}, in that we do not assume a rest frame based on
the invariant mass of the sum of the observed jet constituents.  
 
To demonstrate the separation that these BES variables can provide, we
generate simulated events with the {\sc PYTHIA} Monte Carlo
generator~\cite{PYTHIA} in which we produce a new hypothetical
$Z^\prime$ particle which decays to $t\bar{t}$, $W^+W^-$, or
$b\bar{b}$.  The mass of the $Z^\prime$ is taken to be 3.0 TeV.  This
results in a distribution of the transverse energy of the $Z^\prime$
decay prodicts which is relatively uniform up to about 1.5 TeV.

We then simulate the detector response with the PGS
program~\cite{SHW}, and find jets reconstructed from calorimeter
energy deposits using the anti-$k_T$ algorithm with a cone size of
0.8.  A crude simulation of PF constituents is then achieved by
selecting all generator-level final state particles with $p_T > 3$ GeV
matching a jet, and imposing a uniform 90\% efficiency.  Note that
this results in a list of jet constituents which does not reflect
detector resolution effects or the effect of additional $pp$
(``pileup'') interactions.

\begin{figure*}
  \begin{center}
      \includegraphics[width=\textwidth]{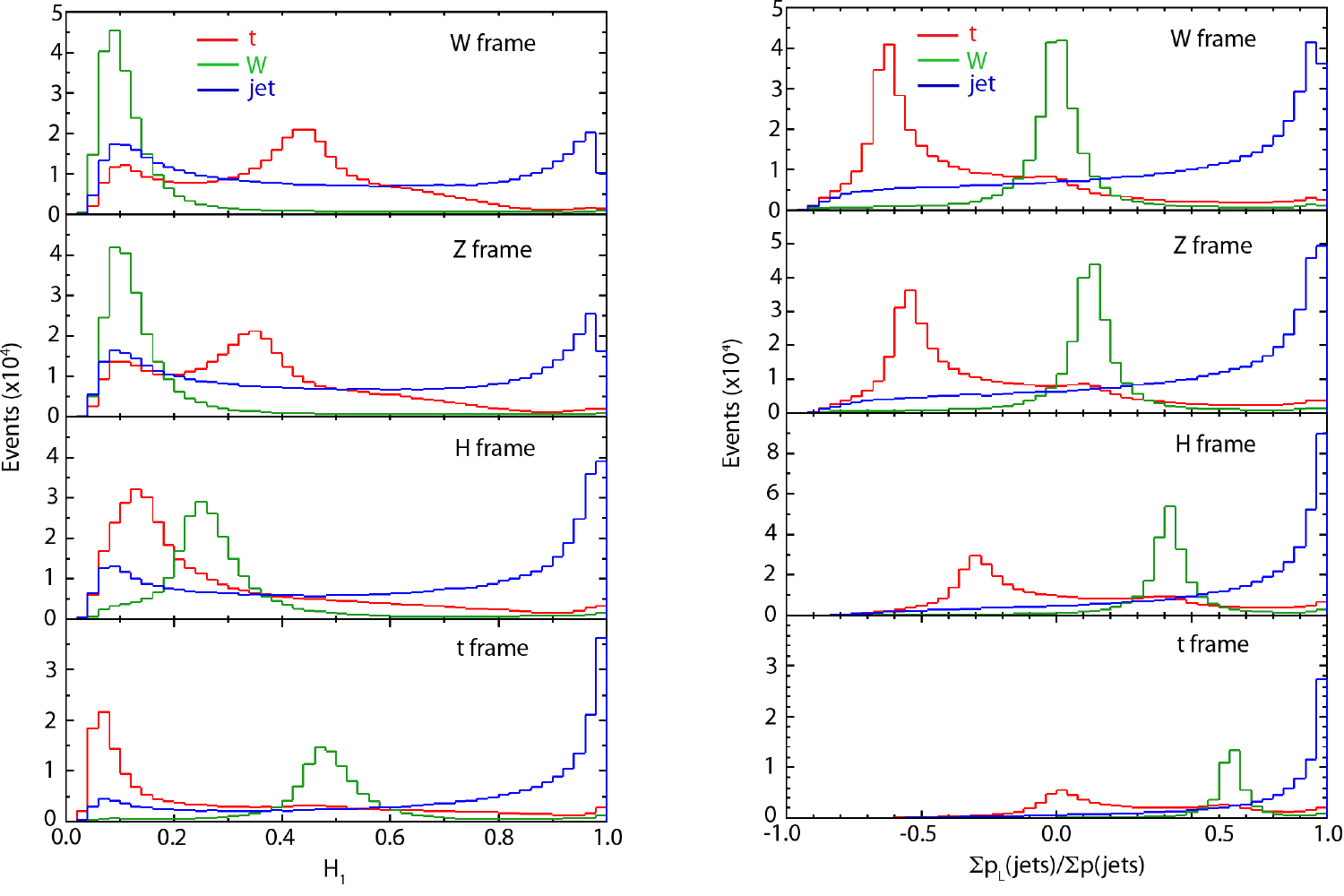}
  \end{center}
  \caption{Distributions of the first Fox-Wolfram moment $H_1$ (left)
    and the ratio of the longitudinal to toal momentum of anti-$k_T$
    jets (right) in various boosted reference frames, for top quarks,
    $W$ bosons, and $b$ quark jets from 3.0 TeV $Z^\prime$ decays.}
  \label{fig-dist}
\end{figure*}

We take the list of jet constituent four-momenta and boost each one
into a reference frame with velocity $\beta=p/E$, where to calculate
$E$ we use the masses of the top quark, Higgs boson, $Z$, and $W$
bosons.  In each frame we then calculate the above-mentioned event
shape quantities.  Since we use the four Fox-Wolfram moments, the
three sphericity tensor eigenvalues, the thrust, and ten quantities
characterizing the jets found in the boosted frame, for a given jet
this results in 4$\times$18=72 different quantities.  Space does not
permit displaying the distributions of all these quantites, but we
show two of them in Figure~\ref{fig-dist}.  These distributions are
based on fat jets found in events in which a 3.0 TeV $Z^\prime$ is
produced and decays to $t\bar{t}$, $WW$, or $b\bar{b}$, as
implemnented in the {\sc PYTHIA} Monte Carlo generator. In these
figures the histograms show how the distributions evolve as one
successively increases the mass used to calculate the boosted frame.

\section{Neural Network BES Discriminator}

With the large number of BES quantities characterizing a given jet, it
is natural to deploy a multivariate technique such as artificial
neural networks (ANN) or boosted decision trees (BDT) to arrive at a
classification of jets as to their source ($t/H/Z/W$ or light quark or
gluon jet).  As an initial test we created training and test samples
using the 3.0 TeV $Z^\prime$ events.  From jets matched to generator
level $t$, $W$, and $b$, required to decay all-hadron final states,
training and test sets are made.  For each event we have 50 input
variables.  From the 72 variables enumerated above we eliminate the
boosted jet $E_T$ and invariant masses for all but the $W$ frame,
since the quantities in the different frames are so highly correlated
from one frame to the next as to be redundant, which is understandable
given their kinematic nature.  This leaves 48 quantities, to which we
add two more: the jet $E_T$ and $|\eta|$.  We arrive at training
samples with 20k events each of $t$, $W$, and $b$ jets, and 15k events
in each of the three test sets.  To avoid the multivariate method
using the $E_T$ as a discriminating variable, we draw training and
test events from the $b$ and $W$ samples according to the distribution
of $E_T$ for jets from top decays.

As a demonstration that this technique can potentially achieve a high
degree of accurate classification, we train a set of feedforward
multilayer perceptrons with various numbers of hidden layers and nodes
in each layer, and two output nodes.  The output nodes are trained to
be (0,0) for $b$ jets, (0,1) for $W$ jets, and (1,0) for $t$ jets.
The ANN is trained using backpropagation with a learning strength
parameter of 0.001, a momentum parameter of 0.5, and updating the
weights and thresholds after every 50 training events.  The learning
strength parameter decays exponentially during training with a time
constant of 500 epochs (complete presentations of the training sets).
Performance of the networks is monitored during training, and we
observe stable performance asymptotically, well before the maximum
number of 1000 training epochs is reached.  We see little variation in
performance as a function of the number of hidden layers and number of
nodes in each layer - in fact we see more variation from one network
to another having the same number of nodes and layers, but starting at
a different random weights and thresholds.  The results presented
here are attained with nets with three hidden layers of 40 nodes each.

\begin{figure}
  \begin{center}
      \includegraphics[width=3.0in,angle=0]{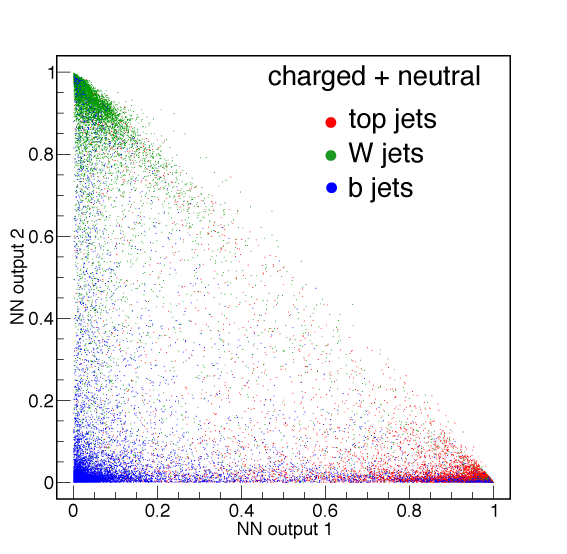}
      \includegraphics[width=3.0in,angle=0]{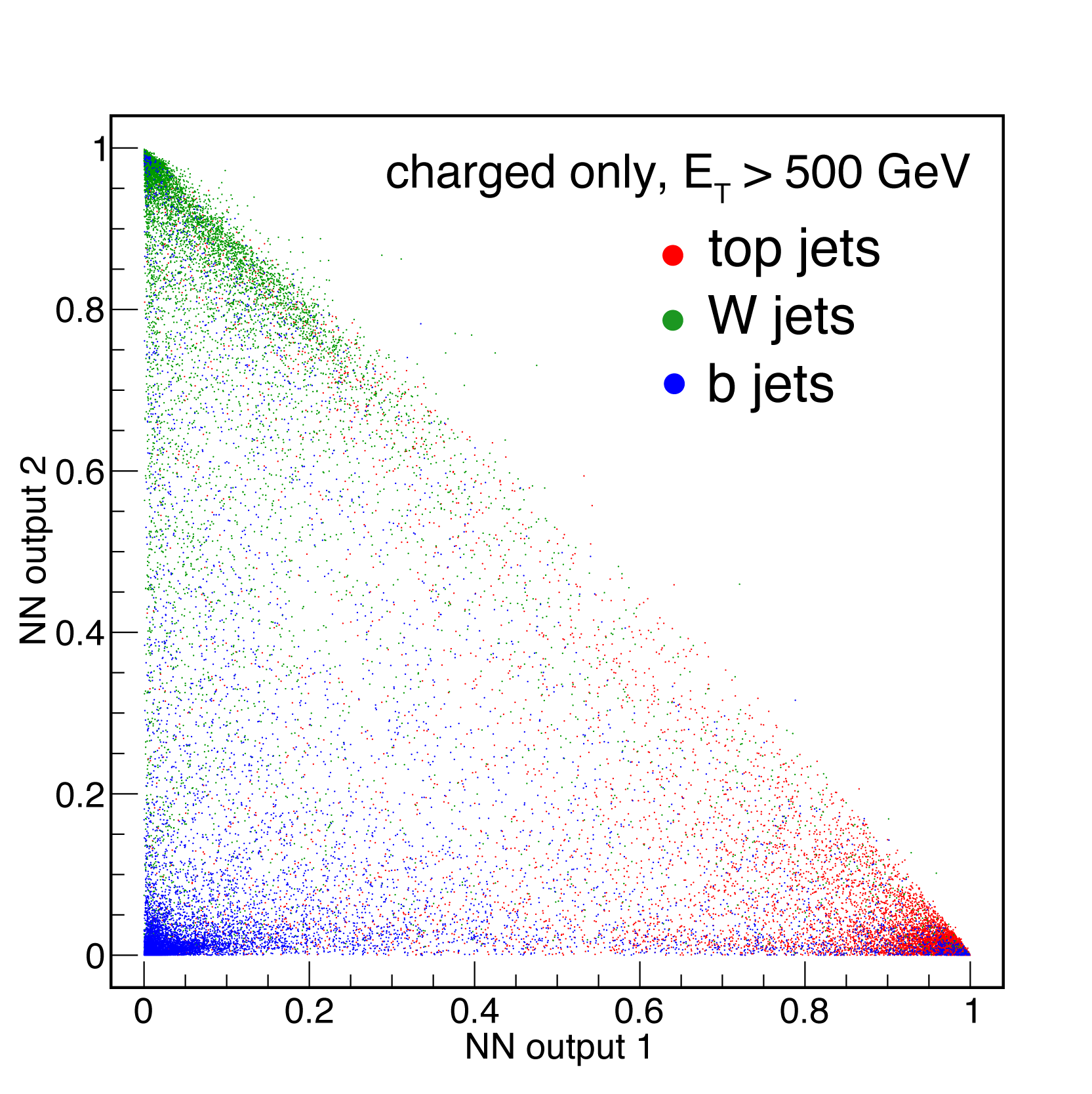}
  \end{center}
  \caption{Distributions of the neural network outputs for test samples
           of jets originating from all hadronic top quark decay, $W$ decay, and 
           $b$ quarks for networks trained with charged and neutral constituents
           (top) and charged conatituents only (bottom).}
  \label{fig-net}
\end{figure}

\begin{figure}
  \begin{center}
      \includegraphics[width=\columnwidth]{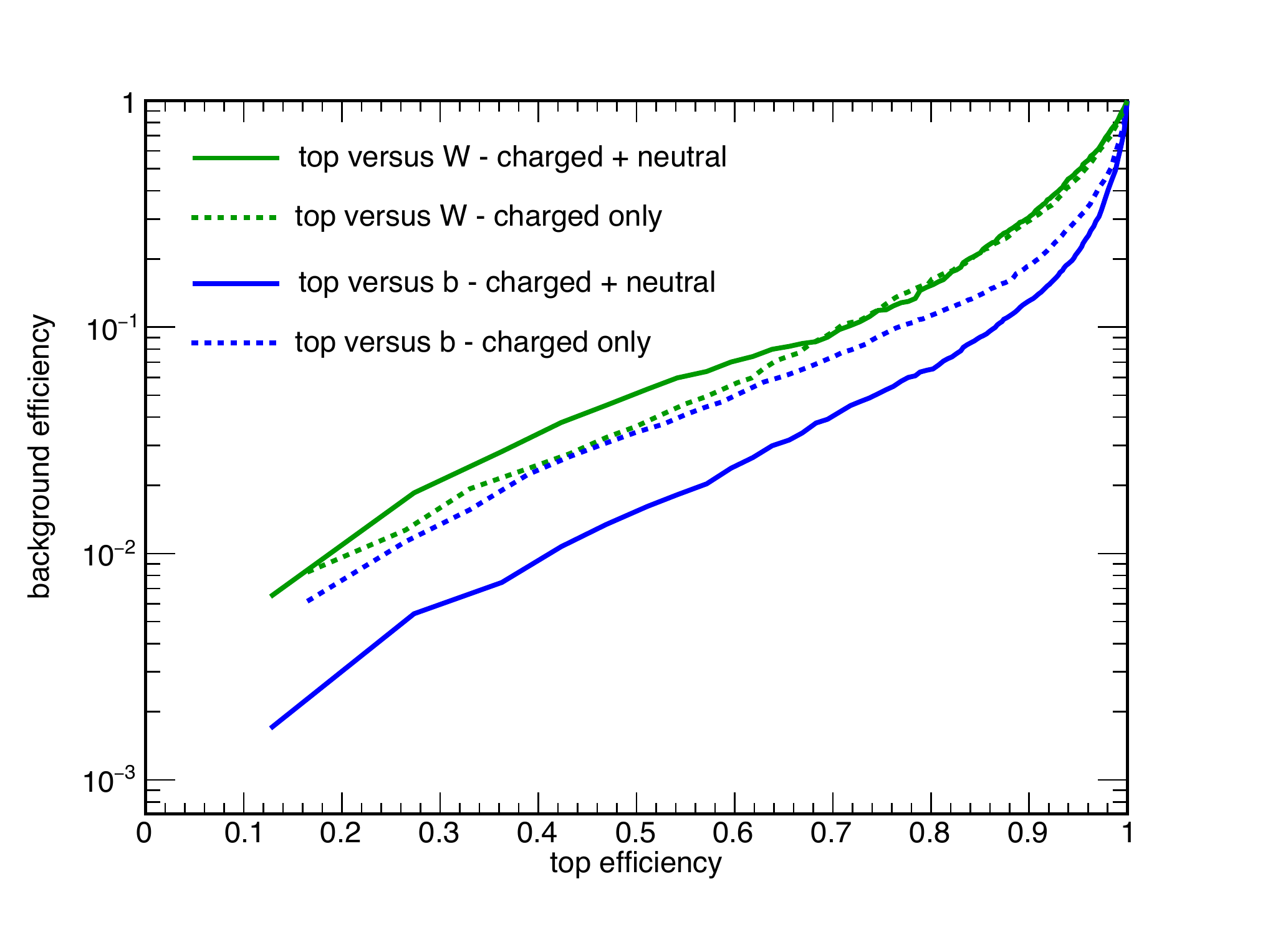}
  \end{center}
  \caption{Efficiency for identification of jets from top decay (horozintal axis)
           versus efficiency for $W$ jets (green) and $b$ jets (blue), showing the
           results using charged and neutral jet constituents (solid curves) and
           using charged constituents only (dashed curves).}
  \label{fig-ROC}
\end{figure}

Figure~\ref{fig-net} illustrates the neural network output
distributions for the three jet types.  The figure shows that there is
a high degree of separation of the three sources.  One can also
characterize the accuracy of classification using a matrix showing how
many true $t$, $W$, and $b$ jets are classified as such.  Using a
simple classification in which the category is taken to be the one
nearest the neural network outputs, we see from Table~\ref{tab-matrix}
that approximately 85\% of jets are classified correctly.  One can
also see the performance in Figure~\ref{fig-ROC}, which show the
efficiency of $t$ and $W$ classification, respectively, versus the
efficiency for the other two categories.  From these curves one can
see, for example, that at an efficiency of 60\% for $t$
identification, one has a background efficiency for $b$ jets of just
over 2\%.

\begin{table}
\begin{center}
  \begin{tabular}{lccc}\hline\hline
Generated:          &     $t$        &    $W$     &    $b$     \\ \hline
Classified as $t$   &    84.2\%      &   3.9\%    &   10.2\%   \\  
Classified as $W$   &     5.6\%      &  86.8\%    &    5.9\%   \\
Classified as $b$   &    10.4\%      &   9.3\%    &   83.9\%   \\ \hline
  \end{tabular}
\end{center}
  \caption{Probabilities for classification of $t$, $W$, and $b$ jets as such 
           using the neural network outputs, training on charged and neutral
           jet constituents.}
  \label{tab-matrix}
\end{table}

While the above results are encouraging, it is clear that in an actual
LHC analysis the performance will likely not attain these levels.
Perhaps the main reason is that in real analyses the identification of
individual jet constituents will not be as efficient or accurate as
assumed in this prototype study.  At very high momentum, in
particular, when there is more confusion between neutral particles, it
may be necessary to rely solely on charged particles. For this reason
we tested the performance of the BES algorithm using charged tracks
only, selecting only those generator-level final state particles which
were charged, having $p_T > 2$ GeV.  Figures \ref{fig-net} and
\ref{fig-ROC} compare the results using charged constituents only to
the results from using both charged and neutral constituents.
Remarkably, the performance for the separation of top quarks from $W$
bosons remains nearly the same, but the performance for separation of
top from $b$ jets is degraded, with background efficiency a factor of
two larger at a top quark efficiency of 60\%.

It is also of interest to examine whether this new algorithm is
capable of distinguishing all-hadronic top quark decays, Higgs boson
decays to $b\bar{b}$, $Z\to q\bar{q}$, and $W\to q\bar{q}^\prime$ when
the decaying particles are higly boosted.  To test this we also used
samples of a 3 TeV mass $Z^\prime$ decaying to $t\bar{t}$, $ZH$,
$W^+W^-$, and $b\bar{b}$.  As before the events are simulated with
{\sc PYTHIA} and passed through the PGS simulation.  Fat jets
(anti-$k_T$, $R=0.8$) are then matched to the decaying boosted objects
and 90\% of the jet constituents with $p_T > 2$ GeV and lying within
$\Delta R < 0.8$ of the jet direction are recorded for the BES
analysis.  We consider here the performance using both charged and 
neutral objects. 

\begin{figure}
  \begin{center}
      \includegraphics[width=3.0in, angle=0]{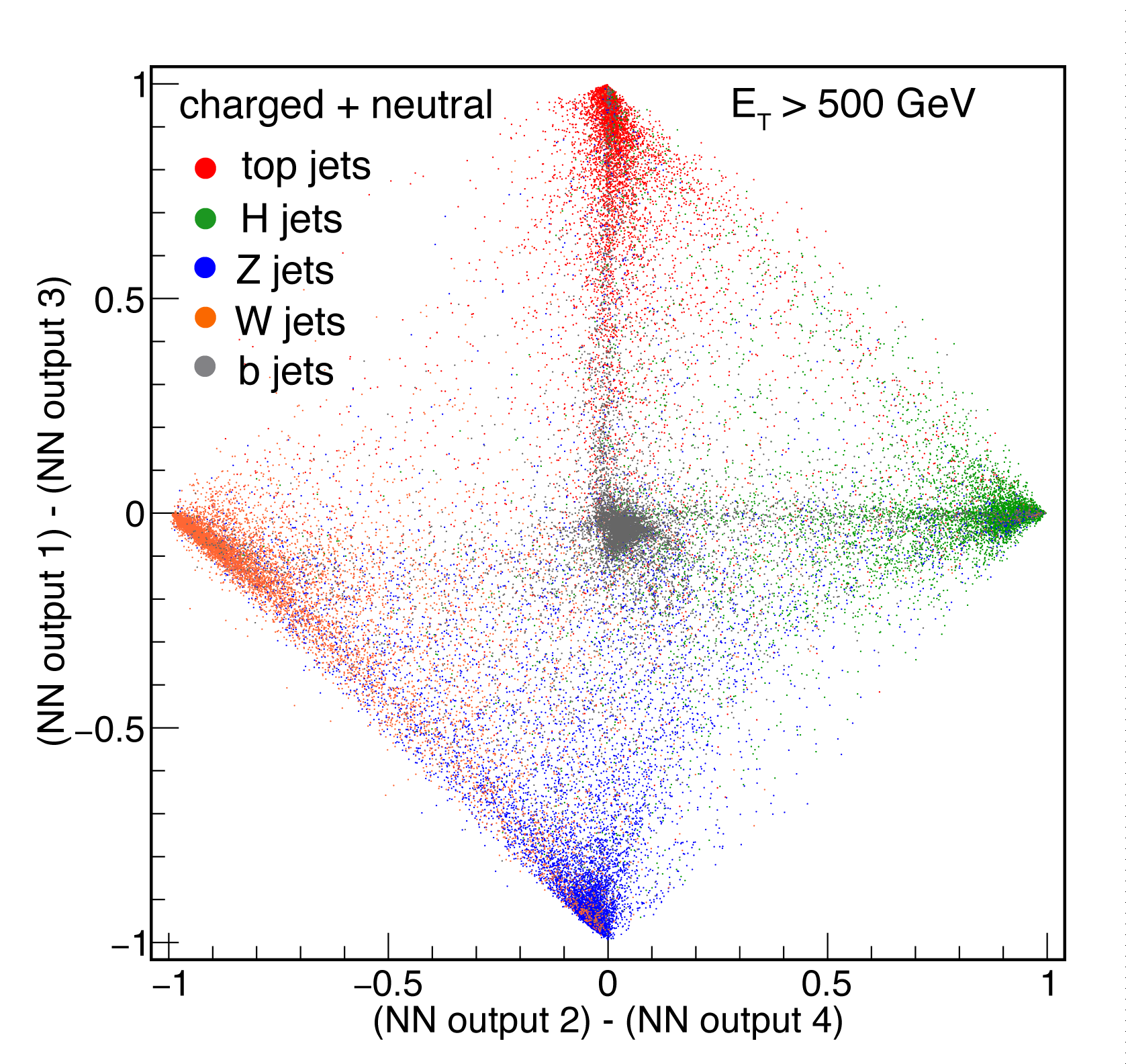}
  \end{center}
  \caption{Distributions of the neural network output differences for 
           a 4-output neural network trained to discriminate jets 
           originating from all-hadronic top quark decay, $H\to b\bar{b}$ 
           decay, $Z\to q\bar{q}$ decay, $W\to q\bar{q}^\prime$ decay, and 
           $b$ quarks for networks trained with charged and neutral constituents.}
  \label{fig-net5}
\end{figure}

\begin{figure}[t]
  \begin{center}
      \includegraphics[angle=90,width=\columnwidth]{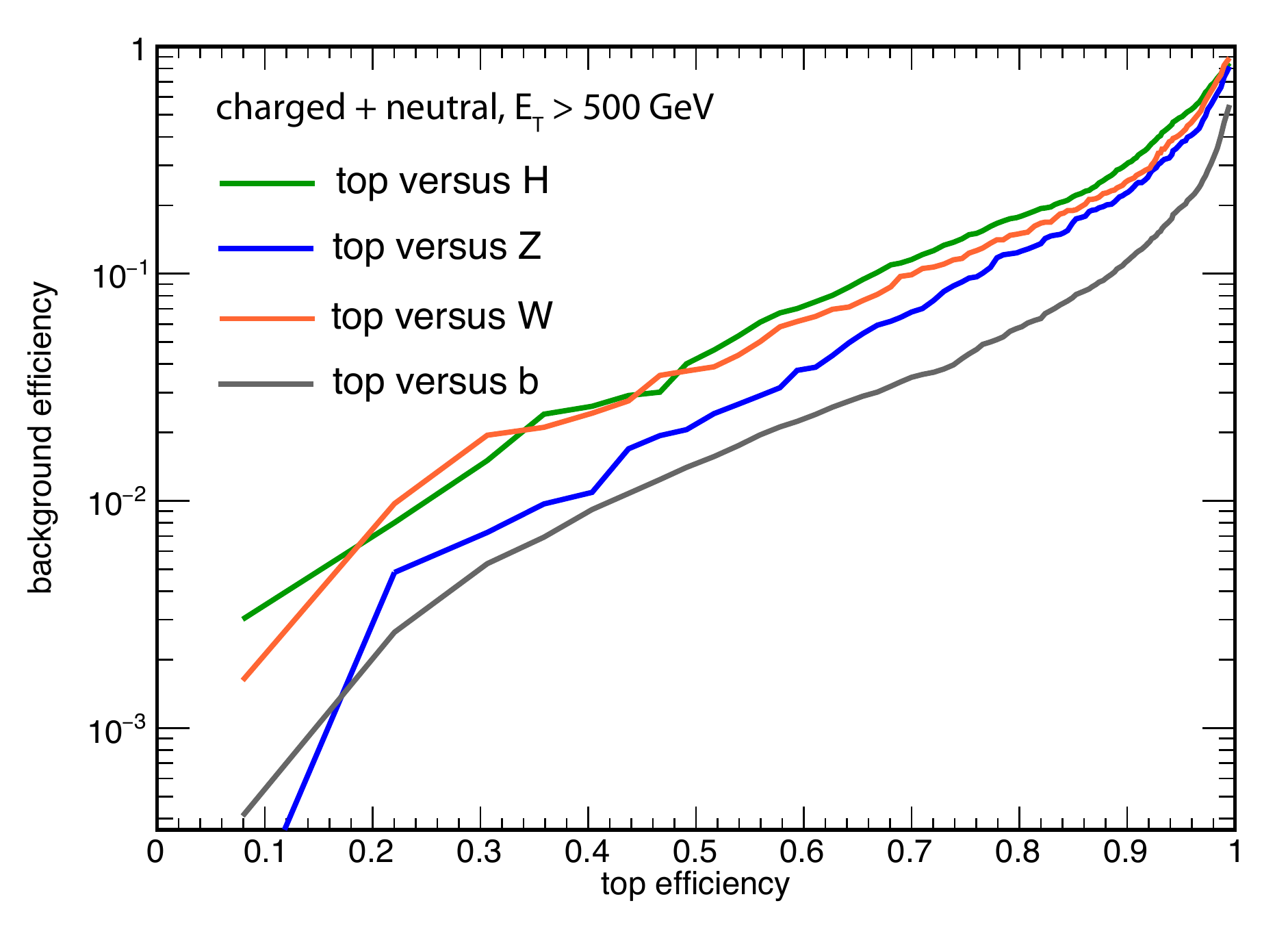}
  \end{center}
  \caption{Efficiency for identification of jets from top decay (horozintal axis)
           versus efficency (vertical axis) for jets originating from all-hadronic decay of $H$,
           $Z$, and $W$ bosons, and $b$ quarks, for a 4-output neural network
           based on BES variables from charged and neutral jet constituents.}
  \label{fig-ROC5}
\end{figure}

We train a 6-layer neural network with the same 50 inputs, 4 layers of
40 hidden nodes each, and an output layer with four outputs.  The 
network is trained to give values of (1,0,0,0) for $t$ jets, (0,1,0,0) for
$H\to b\bar{b}$ jets, (0,,0,1,0) for $Z\to q\bar{q}$, (0,0,0,1) for
$W\to q\bar{q}^\prime$ jets, and (0,0,0,0) for $b$ quark jets.  The 
resulting output, which are points in a 4-dimensional space,  can 
nevertheless be usefully displayed on the two-dimensional plane of the 
difference between the first and third outputs ($y$ axis) and the 
difference between the second and fourth outputs ($x$ axis).  
Figure~\ref{fig-net5} shows these distributions for the five jet 
categories, and illustrates the great potential to discriminate them
well, even $W$ from $Z$ decays.   As an example, Figure~\ref{fig-ROC5}
shows the efficiency for identifying top quark decays as a function of
the efficiency for identifying the other particle types.

The performance of the algorithm may benefit from the
various jet ``pruning'' techniques such as the soft-drop
method~\cite{soft-drop} and {\sc PUPPI}~\cite{PUPPI}, but this is best
studied in the context of a full simulation including all detector
effects and the effets of multiple interactions.  

The theoretical uncertainties on the modeling of rest-frame quantities
were studied in detail elsewhere~\cite{Feige}.  We note that these
can be at least partly overcome by training the neural network using 
background (light quark and gluon) samples derived from observed 
dijet events, where the other jet is anti-tagged as coming from 
$t/H/Z/W$.  

Finally, we emphasize that
the variables arising from the boosted event shapes can be readily
combined with variables used in other approaches, including the
presently used $N_{subjettiness}$~\cite{Thaler} and soft-drop mass,
and other novel approaches such as those based on two-dimensional
imaging of jets~\cite{Baldi}, to arrive at an even more selective
algorithm.  One can expect, hoever, that the the BES variables will
retain their discriminating power up to ultrahigh boosts at which 2D
variables become less discriminating.

\section{Summary}

We have described here a new approach to identifying very high
momentum hadronically decaying top quarks, Higgs bosons, and vector
bosons, based on boosting the individual constituents of jets
sequentially into rest frames corresponding to the hypothetical
particle masses and calcualting in each of those frames a set of event
shape quantities.  These quantities include the Fox-Wolfram moments,
sphericity tensor and its eigenvalues, thrust, and quantities derived
from anti-$k_T$ jets calculated using the boost axis as the ``beam''.
The algorithm is capable of separating jets which strongly overlap
in laboratory momentum space.  In an initial, idealized scenario using
generator-level particles the algorithm shows excellent separation of
hadronic jets arising from the decay of top quarks, Higgs bosons, $Z$
and $W$ bosons and from $b$ quarks.  This level of separation is
largely maintained when using charged jet constituents only, as might
be necessary for ultra-high-energy jets.  The variables arising in
this algorithm can be readily combined with those from other boosted
particle tagging techniques.

\end{document}